\documentclass[9pt,roundtable]{livecoms}

\usepackage{lipsum} 
\usepackage[version=4]{mhchem}
\usepackage{siunitx}
\DeclareSIUnit\Molar{M}
\usepackage[italic]{mathastext}
\graphicspath{{figures/}}
\usepackage[most]{tcolorbox}
\usepackage{bm}\usepackage{stfloats}
\usepackage[numbers]{natbib}
\usepackage[utf8]{inputenc}

\newcommand{\authorblock}[1]{\vskip\smallskipamount
\fcolorbox{LiveCoMSLightBlue!10}{LiveCoMSLightBlue!10}{
\begin{minipage}{\linewidth}
  \textbf{Authors:}\\
  #1
\end{minipage}
}
\vskip\baselineskip
}
\newcommand{\kBT}{k_{\mathrm B}T}
\newcommand{\Kzero}{K^{(0)}}
\newcommand{\Kb}{K^{(b)}}
\newcommand{\pizero}{\pi^{(0)}}
\newcommand{\bd}{\bm b}
\newcommand{\bg}{\bd^{*}_{\gamma}}
\newcommand{\bmfpt}{\bd^{*}_{\rm MFPT}}
\newcommand{\tauAB}{\tau_{AB}}
\newcommand{\xbf}{\mathbf{x}}

\newcommand{\versionnumber}{1.0}  
\newcommand{\githubrepository}{\url{https://github.com/meyresearch/MSM-Roundtable}}  

\title{LiveCoMS Roundtable: Discussing Markov state models after two decades of development [Article v\versionnumber]}

\author[1*]{Daniel M. Zuckerman}
\author[2*]{Gregory R. Bowman}
\author[3*]{Amedeo Caflisch}
\author[4*]{Aaron R. Dinner}
\author[5]{Ahmed Fadhluddin}
\author[3]{Giuseppe Giangreco}
\author[6*]{Xuhui Huang}
\author[7*]{Bettina G.~Keller}
\author[8,9]{Stefan Klus}
\author[10]{Feliks N\"{u}ske}
\author[6]{Michael O'Connor}
\author[5*]{Edina Rosta}
\author[11*]{Christof Schuette}
\author[12*]{Gerhard Stock}
\author[13*]{Vincent Voelz}
\author[14]{Jonathan Weare}
\author[15*]{Antonia S.J.S. Mey}
\affil[1]{Department of Biomedical Engineering, Oregon Health and Science University, USA}
\affil[2]{Departments of Biochemistry \& Biophysics and Bioengineering, University of Pennsylvania, USA}
\affil[3]{Department of Biochemistry, University of Zurich, Switzerland}
\affil[4]{Department of Chemistry and James Franck Institute, University of Chicago, Chicago, IL 60637, USA}
\affil[5]{Department of Physics and Astronomy, University College London, UK}
\affil[6]{Department of Chemistry, Theoretical Chemistry Institute, University of Wisconsin-Madison, Madison, Wisconsin 53706, USA}
\affil[7]{Department of Biology, Chemistry, Pharmacy, Freie Universit\"at Berlin, Germany}
\affil[8]{Maxwell Institute for Mathematical Sciences, University of Edinburgh and Heriot–Watt University, Edinburgh, United Kingdom}
\affil[9] {Department of Mathematics, Heriot–Watt University, Edinburgh, United Kingdom}
\affil[10]{Max-Planck-Institute for Dynamics of Complex Technical Systems, Magdeburg, Germany}
\affil[11]{Freie Universit\"{a}t Berlin, Institute of Mathematics and Computer Science, Berlin, and Zuse Institute Berlin, Germany}
\affil[12]{Department of Physics, University of Freiburg, Germany}
\affil[13]{Department of Chemistry, Temple University, USA}
\affil[14]{Courant Institute of Mathematical Sciences, New York University, New York, NY 10012, USA}
\affil[15]{EaStCHEM School of Chemistry, University of Edinburgh, EH10 3FJ, United Kingdom}

\corr{zuckermd@ohsu.edu}{DMZ}  
\corr{antonia.mey@ed.ac.uk}{ASJSM}

\orcid{Gregory R. Bowman}{0000-0002-2083-4892}
\orcid{Amedeo Caflisch}{0000-0002-2317-6792}
\orcid{Aaron R. Dinner}{0000-0001-8328-6427}
\orcid{Giuseppe Giangreco}{0009-0000-8297-632X}
\orcid{Xuhui Huang}{0000-0002-7119-9358}
\orcid{Bettina G.~Keller}{0000-0002-7051-0888}
\orcid{Stefan Klus}{0000-0002-9672-3806}
\orcid{Antonia S.J.S. Mey}{0000-0001-7512-5252}
\orcid{Feliks Nüske}{0000-0003-2444-7889}
\orcid{Michael S. O'Connor}{0000-0003-3551-7248}
\orcid{Edina Rosta}{0000-0002-9823-4766}
\orcid{Christof Schuette}{0000-0001-5232-2683}
\orcid{Gerhard Stock}{0000-0002-3302-3044}
\orcid{Vincent Voelz}{0000-0002-1054-2124}
\orcid{Jonathan Weare}{0000-0001-8745-1821}
\orcid{Daniel M. Zuckerman}{0000-0001-7662-2031}

\presentadd[\authfn{3}]{Department of Biochemistry, University of Zurich, Switzerland}

\blurb{This LiveCoMS document is maintained online on GitHub at \githubrepository; to provide feedback, suggestions, or help improve it, please visit the GitHub repository and participate via the issue tracker.}

\pubDOI{10.XXXX/YYYYYYY}
\pubvolume{<volume>}
\pubissue{<issue>}
\pubyear{<year>}
\articlenum{<number>}
\datereceived{Day Month Year}
\dateaccepted{Day Month Year}

\begin{document}
    
\begin{frontmatter}
\maketitle

\begin{abstract}
Markov state models (MSMs) play a central role in the analysis of molecular dynamics (MD) simulations for biomolecules, and MSM methodologies have been under continual development since their introduction to the field in the early 2000s.  MSMs take MD trajectory data as input and produce a more compact, i.e., coarse-grained, and interpretable representation of the system in terms of conformational states and quantitative observables such as the equilibrium populations of states and the timescales of interconverting between states.  Numerous strategies and considerable mathematical formalism have been developed to improve coarse-graining in both time and space to enable the most reliable representation of the underlying biomolecular dynamics.  Nevertheless, there remains considerable discussion in the literature regarding best practices and MSM limitations, as well as proposals to abandon or extend the first-order Markovian framework in common use.  This Roundtable article gathers together diverse opinions from $> 10$ authors active in MSM development, and by design, each author or author group offers a separate contribution.  No consensus view is provided.  Instead, the article is meant to be a guide for readers wishing to be exposed to a wide set of expert perspectives.



\end{abstract}

\end{frontmatter}



\section*{Introduction}
\authorblock{Daniel M. Zuckerman and Antonia S.J.S. Mey}
Markov state models (MSMs) 
have provided 
a mathematically founded framework to analyze molecular dynamics (MD)-based data starting in the late 1990s and early 2000s, when physics-based simulation methods were used to understand conformational dynamics of molecules and, more ambitiously, to address the protein folding problem~\cite{dill2012protein-folding}. In the seminal paper by Schütte et al.~\cite{schutte1999direct}, transfer operator theory and Markov chains for a state space analysis were linked for the first time to the analysis of conformational dynamics of small molecules. Rao and Caflisch brought this to the world of protein folding by using graph-theoretic approaches to understand protein folding kinetics~\cite{rao2004protein}, and Swope et al. performed the first systematic construction of MSMs from MD data~\cite{swope2004msm-one,swope2004msm-two}. Since then, MSMs have been widely used to analyze biomolecular MD simulations --- from protein folding~\cite{voelz2010molecular} to protein-ligand binding~\cite{silva2011binding,plattner2015protein} to RNA dynamics~\cite{pinamonti2016predicting} --- but also applied to entirely different fields, such as climate science, to extract equilibrium and kinetic properties from simulations~\cite{finkel2023revealing}. 

Over the last 20 years, mathematical, hardware, and software development for the generation of MD data~\cite{beck2024all-atom} and MSM building~\cite{bowman2009using,harrigan2017msmbuilder,scherer2015pyemma,senne2012emma,porter2019enspara, hoffmann2021deeptime,turri2026kooplearn} has seen the field grow enormously. Yet with a whole zoo of methods available, it can seem overwhelming to choose the best strategies to extract equilibrium and kinetic information from simulations~\cite{pande2018review}. 
Several recent works have critically examined MSM approaches and variants \cite{vitalis2019removal,suarez2020what,he2021critical,grubmuller2023uncertainties,nagel2023benchmark,mey2024optimize}, providing valuable if incomplete guidance.
In addition, excellent resources exist in terms of textbooks~\cite{bowman2014introduction} and review articles~\cite{prinz2011optimal,pande2018review,chodera2014markov}, but MSM building can still be something of an art, particularly with recent advances incorporating ML, such as VAMPnets~\cite{mardt2018vampnets}.

The lack of consensus among experts, along with the goal of exposing beginners and practitioners alike to the state of the art --- including criticisms and alternatives --- motivate this pilot Roundtable article.  Each author or author group has written separately to provide their own perspective, but authors had the opportunity to update their contributions based on the other perspectives.  As with all articles in the journal, this one will be [has been] subjected to peer review.  The end result, we hope, will deepen readers' understanding of key issues of concern when constructing an MSM (or alternative) from the data-generation phase to model interpretation.

A wide range of insightful viewpoints is presented. Bowman emphasizes a key goal of applying MSMs for quantitative comparison to experimental data.  Caflisch and Giangreco raise the issue of sampling biomolecular simulations to convergence and discuss strategies to work with unconverged data. Dinner and Weare discuss dynamical Galerkin approximations (DGA), a generalization of MSMs for computing kinetic observables. O'Connor and Huang emphasize the role of memory in extending MSMs, as well as integrating optimized features and states for characterizing transition states.  Keller presents a broad view of MSMs framed as statistical learning, noting error sources and the role of memory even in optimized spaces.  Klus, Mey, and Nüske discuss extensions to traditional MSMs by approximating the Koopman operator through different strategies.  Rosta and Fadhluddin describe a biasing approach for enhanced sampling that is optimized within an MSM framework based on the mean first-passage time.  Schütte takes a broad theoretical view and describes MSMs as one, among many, approaches for coarse-graining dynamics, and argues for the importance of optimized collective variables.  Stock emphasizes the feature selection step necessary for all modeling, focusing on a community detection approach rooted in pairwise distances.  Voelz turns a critical eye to MSM machine learning based on the variation approach to Markov processes (VAMP) and notes challenges for MSMs in computing equilibrium populations.  Zuckerman describes an analysis based on two transition matrices that abandons the Markov assumption and can provide unbiased estimates of key observables.

\paragraph{Quick MSM Essentials}

For the novice reader, the following provides a brief introduction to the ideas
behind Markov modeling;
see also the glossary of terms and symbols used throughout this article. 
Generally, the goal of a
Markov state model (MSM) is to capture biologically or chemically relevant
conformational changes from simulation data in terms of coarse-grained states (``clusters'') in a discrete-time description.
On this basis, the MSM provides a mathematical framework
to extract the timescales and equilibrium properties of the system of study.

MSMs are trained from trajectory data and analyzed using stochastic matrices \cite{chodera2014markov}.
Typically, a molecular dynamics (MD) trajectory is featurized (for example, into
distances or dihedral angles) and, often after a dimensionality reduction onto a
few collective variables (CVs) or a reaction coordinate (RC) $\xi(\mathbf{x})$, the data is
discretized into states, from which a transition matrix $\mathbf{T}$ (with
elements $T_{ij}$) is built from the trajectory data over a chosen lag time
$\tau$. Each element is the conditional transition probability from state $i$ to $j$ in time $\tau$. The eigenvalue decomposition of this matrix provides eigenvectors that
give information on the slow processes in the system, while the corresponding
eigenvalues $\lambda_i$ give the associated implied timescales (ITS),
$t_i(\tau)$; the leading eigenvector is the stationary distribution $\pi$, which
encodes the equilibrium populations. 

Beyond timescales and populations,
transition path theory (TPT) provides observables that describe how a transition
occurs between a chosen reactant set $A$ and product set $B$, such as the mean
first-passage time (MFPT) $\tau_{AB}$ and the committor $q(\mathbf{x})$.
Traditionally the states are discrete and $\mathbf{T}$ is a stochastic Markov
matrix, however there are now many variants where this is no longer the case (see Perspectives below):
the matrix can be generalized to a transfer operator $\mathcal{T}(\tau)$ or a
Koopman operator $\mathcal{K}^\tau$ (or its generator $\mathcal{L}$), estimated
for example by extended dynamic mode decomposition (EDMD), memory can be
reintroduced through a generalized master equation (GME), and the data may even
come from a nonequilibrium steady state (NESS) rather than equilibrium.
Validation of the model can be done through a long-timescale test called the
Chapman--Kolmogorov test. Modeling choices, such as the dimensionality
reduction (often time-lagged independent component analysis, tICA), the number
of states, and the lag time, can then be tried by optimizing a variational
score (e.g., VAC or VAMP), for instance to build a model that best captures the
slow processes in the system.

\newpage
\section{Glossary}
\label{sec:def}
\begin{tcolorbox}[title=List of Abbreviations, colback=green!10!white, breakable]
\small\raggedright
MSM --- Markov state model \\
MD --- molecular dynamics \\
MFPT --- mean first-passage time \\
ITS --- implied timescale(s) \\
CV --- collective variable \\
RC --- reaction coordinate \\
TPT --- transition path theory \\
tICA --- time-lagged independent component analysis \\
VAC --- variational approach to conformational dynamics \\
VAMP --- variational approach to Markov processes \\
EDMD --- extended dynamic mode decomposition \\
GME --- generalized master equation \\
NESS --- nonequilibrium steady state \\
\end{tcolorbox}

\begin{tcolorbox}[title=List of Symbols, colback=green!10!white, breakable]
\small\raggedright
$\mathbf{x},\ \mathbf{x}_t$ --- point in phase space (of all system coordinates), at time $t$ \\
$\tau$ --- lag time \\
$p_t(\mathbf{x}_t)$ --- distribution of phase points at time $t$ \\
$p^\mathrm{prop}(\mathbf{x}_{t+\tau}\,|\,\mathbf{x}_t)$ --- ``microscopic'' transition probability \\
$\mathbf{T},\ T_{ij}$ --- transition matrix (row-stochastic) and its elements \\
$\pi$ --- stationary distribution / vector \\
$\lambda_i$ --- $i$-th eigenvalue of the transition matrix \\
$t_i(\tau)=-\tau/\log\lambda_i(\tau)$ --- implied timescale \\
$A,\,B$ --- reactant/source set and product/target set \\
$\tau_{AB}$ --- mean first-passage time from $A$ to $B$ \\
$q(\mathbf{x})$ --- committor (probability of reaching $B$ before $A$) \\
$\mathcal{T}(\tau)$ --- transfer (propagator) operator \\
$\mathcal{K}^{\tau}$ --- Koopman operator \\
$\mathcal{L}$ --- generator of the dynamics \\
$\xi(\mathbf{x})$ --- collective variable / reduced coordinate \\
\end{tcolorbox}


\setcounter{section}{0}  
\newpage
\section{Perspective \thesection: Fine-grained models for quantitative prediction of experiments}
\authorblock{Gregory R. Bowman*}
There are a lot of physically-sound MSMs for a given biomolecular system~\cite{bowman2014introduction}, so I am not concerned by the heterogeneity in how different scientists build MSMs. Much of this heterogeneity reflects differences in the problems being addressed and how a given scientist approaches trade-offs between different approaches to MSM construction.

One of the most significant trade-offs that must be addressed when choosing how to build an MSM is between interpret ability and quantitative prediction of experimental observables.

My lab and I typically address this trade-off by using fine-grained models built via RMSD-based clustering for quantitative analysis, paired with coarse-grained models for interpretation and visualization~\cite{bowman2009using}. All our code for building MSMs is available on GitHub, notably through our enspara software \url{https://github.com/bowman-lab/enspara}~\cite{porter2019enspara}. 

Fine-grained models (typically having thousands to tens-of-thousands of small states) are excellent for making a quantitative connection with experiments, as such a connection generally benefits from having high spatial and temporal resolution. Experiments do not magically detect large free energy barriers and report on the kinetics of crossing them or the equilibrium probabilities of falling on each side. They report on observables that stem from the details of the protein structure, which may or may not correlate with the dominant basins in the underlying free energy landscape. If you want to make a quantitative connection with experiments, then a generally reliable way is to have states where conformations within a state are so similar that it is reasonable to assume they give rise to the same observable signal, whatever experimental observable you choose. In contrast, MSMs with tens of states may be easier to manage but the great conformational heterogeneity within each state often makes quantitative predictions of experiments challenging.

To build fine-grained models, we like to do an RMSD-based clustering (typically based on backbone heavy atoms and C$\beta$ atoms) using a hybrid k-centers/k-medoids algorithm~\cite{beauchamp2011msmbuilder2, porter2019enspara}. This method uses a k-center algorithm to get roughly evenly dispersed cluster centers. More precisely, it gives states with approximately the same radius. Then, a few rounds of k-medoids updates move the cluster center to the most densely populated part of the state. This update helps ensure that the cluster center is representative of the conformations in the state. Without it, k-centers explicitly favors choosing outliers as cluster centers, which are less likely to be representative of what most of the conformations in the state look like.

In this approach, one can set how big they want states to be (in terms of their radius in RMSD) instead of having to choose the number of states a priori. At the end, you know that all the conformations within a state are within that RMSD from the exemplar, which is a valuable bit of intuition to have about the model as you reason about it.

RMSD-based clustering should always work in the limit of infinite data. Mathematically, dividing a large basin into multiple smaller states doesn’t hurt the global properties of the model and helps with kinetics~\cite{sarich2010approximation}.

Of course, there are downsides. The more states you have, the more data you need to parameterize the transition probability matrix. Large datasets from Folding@home~\cite{voelz2023folding, meller2024physicsbased} or large supercomputer allocations are one way to obtain sufficient data. Adaptive sampling also provides a means to gather the most useful data with more limited compute resources~\cite{zimmerman_fast_2015,zimmerman2018choice}. It is also hard to look at a model with $50,000$ states, either to interpret it or to make a clean figure. And sometimes you need a truly astronomical number of states for this approach to work. Intrinsically disordered proteins are a good example. That is one case where I recommend people do not start by trying an RMSD-based clustering~\cite{stuchelle-brereton2023Apolipoprotein}.

My lab conducts an initial check on model validity and selects the lag time for the model based on the implied timescales test~\cite{swope2004msm-one,swope2004msm-two}. We are not super picky, though; mostly we use this test to weed out really bad models. Our preferred method for calculating transition probabilities from transition counts is to add a pseudocount of $\frac{1}{N}$ (where $N$ is the number of states) to every entry of the transition count matrix and then to normalize every row~\cite{zimmerman2018choice}. This approach is typically faster and more robust than alternatives (e.g. the popular maximum likelihood estimator~\cite{pande2009progress}).

The real test of a model is quantitative comparison to experimental data. The fewer assumptions or approximations one has to make to calculate an experimental observable, the better. Some of our favorites are NMR order parameters~\cite{bowman2016accurately}, single-molecule FRET distributions~\cite{miller2024accounting}, and populations of open cryptic pockets~\cite{zhang2026well}.

For interpretation and visualization, there are a number of methods for coarse-graining a fine-grained MSM to obtain a more manageable model with tens of states~\cite{deuflhard2000identification,deuflhardRobustPerronCluster2005b,bowman2012improved}. Alternatively, other methods for building MSMs are discussed throughout this article. First-order MSMs are convenient for interpretation but including memory effects can yield better quantitative results, especially for kinetics, which are more sensitive than thermodynamics.

\newpage

\section{Perspective \thesection: Influence of limited sampling on MSM and other methods}
\authorblock{Giuseppe Giangreco and Amedeo Caflisch*}


\subsection{Detailed balance should not be imposed to unconverged sampling}
There are two major challenges with MSMs of simulation data of biological macromolecules. First, they require a grouping of MD snapshots (configurations) into clusters that should obey the Markov property. Second, kinetics and thermodynamics determined by the transition matrix and stationary distribution of a MSM are very sensitive to the MD sampling, which for biological macromolecules is almost always unconverged due to the finite length of MD run(s) and initial state bias \cite{scalco2011equilibrium,vitalis2019removal}.
The bias originates from the structures selected for starting the simulation(s) which are usually very far from a Boltzmann distribution.
Thus, the count matrix is usually noisy and does not obey detailed balance \cite{pande2009progress,pande2010everything}.
Frequently, detailed balance has been imposed by symmetrization of the count matrix or restrained Maximum likelihood \cite{prinz2011markov,trendelkampschroer2015estimation}. Imposing detailed balance can introduce large errors in the kinetics and steady state \cite{scalco2011equilibrium,vitalis2019removal}. Moreover, if both the stationary distribution and the transition matrix are inferred from unconverged sampling, the steady state is usually close to the sampling distribution \cite{scalco2011equilibrium}. Therefore, the resulting MSM is merely a depiction of the data in a coarse-grained (cluster) representation. It does not provide information on sampling convergence.

As an alternative to imposing detailed balance to unconverged sampling, Scalco and one of us  proposed the use of the largest ergodic component (LEC) for building a MSM \cite{scalco2011equilibrium}. The essential idea is to eliminate poorly sampled regions of the configurational space that originate from the non-Boltzmann distribution of starting structures and/or the finite length of simulations. The MSM with imposed detailed balance on the whole sampling and the MSM restricted to the LEC were assessed on distributed computing sampling of a simplified-sequence variant of protein G \cite{guarnera2009how}. Comparative analysis was possible due to the availability of a gold standard, which consisted of a very long trajectory with multiple folding and unfolding events (at 330~K with implicit solvent \cite{haberthuer2009facts}). The stationary distribution and kinetics obtained by the LEC-MSM of distributed computing showed free energy profiles and mean first passage times significantly closer to the gold standard than the MSM of the whole sampling \cite{scalco2011equilibrium}.
It also emerged that imposing detailed balance is helpful only for the LEC. Otherwise, imposing detailed balance has an erratic behavior as it usually introduces short cuts and thus erroneously accelerate the kinetics but it can also artificially slow down the kinetics when starting from transition state structures \cite{scalco2011equilibrium}.
\subsection{Methods that do not require clustering}
MSMs require a priori clustering. One major issue with any clustering procedure is that a fine-grained grouping results in a very sparse transition matrix whereas a coarse grouping of configurations does not preserve the local kinetics. We mention here three methods for analyzing kinetics and states that are not influenced by clustering: the optimized reaction coordinate (RC) \cite{krivov2013reaction}, SAPPHIRE analysis \cite{blochliger2014resolution}, and  RiteWeight \cite{kania2024riteweight}.
Krivov and Karplus proposed the cut-based free-energy profile for the analysis of protein folding \cite{krivov2004hidden,krivov2006onedimensional}. It originally required clustering. Recently, Banushkina and Krivov have introduced a procedure for the unsupervised determination of an optimized RC by minimization of the total squared displacement, requiring minimal intervention \cite{krivov2015nonparametric,krivov2021nonparametric}.

 SAPPHIRE (States And Pathways Projected with HIgh REsolution) analysis is a scalable procedure developed for ordering the configurations from MD sampling or any dynamical system evolving continuously in time \cite{blochliger2014resolution}. The procedure requires only a geometrical distance between MD snapshots as input \cite{caflisch2013scalable}. The essential idea is the use of geometric similarity as a proxy of kinetic vicinity for the construction of the free-energy projection.  The resulting profile is supplemented by a trace of the temporal evolution of the system which illustrates the sequence of events. Furthermore, the SAPPHIRE profile can be annotated with multiple types of (geometric) variables. The resulting plot provides a comprehensive picture of the thermodynamics and kinetics of complex systems at the resolution of individual MD snapshots \cite{blochliger2014resolution}.

RiteWeight is a recently developed iterative procedure for the calculation of the stationary distribution from unconverged simulation data \cite{kania2026randomized}. It estimates the weights of individual trajectory segments of any length and thus it is not influenced by the choice of lag time. The essential idea is the use of random clustering at each reweighting iteration. The repeated random clustering eliminates the discretization error inherent in any grouping procedure. RiteWeight was successful in generating correct steady states from biased distributions of the Trp-cage for which there exists a long simulation as the gold standard \cite{kania2026randomized}.

We propose to analyze the sampling of (poly)peptide folding or aggregation, which is usually unconverged, by MSM and at least one method that does not depend on clustering \cite{krivov2013reaction,blochliger2014resolution,kania2026randomized}.
SAPPHIRE analysis can be used for determining states (clusters) for MSMs \cite{caflisch2020sapphire}, and the optimized RC is useful for the reweighting of adaptive sampling \cite{widmer2023optimized}.

\newpage
\newpage
\section{Perspective \thesection: Computing kinetic statistics with MSMs}
\authorblock{Aaron R. Dinner* and Jonathan Weare}

Markov state models (MSMs) can be used to compute long-time kinetic statistics that reveal transition mechanisms. Such statistics are key elements of transition path theory (TPT) \cite{vanden-eijnden2006transition,metzner2006illustration,e2010transition-path} and include the mean first-passage time from individual microscopic states and the committor $q({\bf x})$, the probability of reaching the product state $B$ from $\bf x$ before reaching the reactant state $A$ \cite{du1998transition,ma2005automatic}. However, early attempts to use MSMs for this purpose yielded results with large uncertainties, with values grouped into a few bins \cite{noe2009constructing,malmstrom2014application,banerjee2014transition}. Instead, MSMs have largely been used to learn relaxation times, related to the eigenvalues of the transition matrix, $\mathbf{T}$ \cite{bowman2014introduction}. One of the main issues limiting the accuracy of statistics produced by MSMs is that the boundary conditions associated with the reactant and product states were not treated correctly \cite{dinner2019galerkin,dinner2021koopman,lorpaiboon2026exact,thiede2025error}. Traditionally, $\mathbf{T}$ is built first and then used to solve the equations of TPT; not accounting for boundary conditions in $\mathbf{T}$ introduces errors. Dynamical Galerkin approximation (DGA) \cite{dinner2019galerkin,dinner2021koopman} is a generalization of MSMs that corrects this issue.

$\mathbf{T}$ can be viewed as the realization of the transition operator $\cal T$ in a 
basis (as discussed below). 
DGA shifts the focus to solving equations of the {\it stopped} transition operator ${\cal S}^\tau f(\mathbf{x})  = \mathbf{E}[f(\mathbf{x}_{\min(\tau, \tau^*)})|\mathbf{x}_0 = x]$, where $\mathbf{E}$ is an expectation and
$\tau^*  = \min\{ t \geq 0|\mathbf{x}_t \in A\cup B\}$ is the time of first entry into $A$ or $B$ \cite{dinner2021koopman}.  The operator is stopped in the sense that points of trajectories subsequent to $\tau^*$ do not contribute to statistics. The committor satisfies ${\cal S}^\tau q({\bf x}) = q({\bf x})$, with $q({\bf x}) = 0$ for ${\bf x}\in A$ and $q({\bf x}) = 1$  for ${\bf x}\in B$ for any lag time $\tau$. For a generic statistic $u({\bf x})$, the equation takes the form ${\cal S}^\tau u({\bf x}) - u({\bf x}) = \int_0^\tau a({\bf x}_t)dt$ for ${\bf x}\notin A\cup B$, with boundary condition $u = b({\bf x})$ for ${\bf x}\in A\cup B$.

DGA solves this equation by expanding the statistic in a basis: $u({\bf x}) = \phi_0({\bf x}) + \sum_{i=1} c_i \phi_i({\bf x})$. Here, $\phi_0({\bf x})$ accounts for the boundary conditions, and $\phi_i({\bf x})$ for $i > 0$ vanish on the reactant and product state boundaries; $c_i$ are coefficients. Substituting the expansion into the equation and applying standard manipulations reduces the problem to one solvable by linear algebra, with matrix elements of the form $\int \phi_i({\bf x}) {\cal S}^\tau\phi_j({\bf x})\mu({\bf x})\, d{\bf x}$; $\mu({\bf x})$ is a sample distribution \cite{dinner2021koopman}.

An explicit form for $\cal S^\tau$ is not available when 
$\tau$ is finite. However, 
the matrix elements above
can be approximated as a sample average over trajectories of length $\tau$: $(1/N)\sum_{k=1}^N \phi_i({\bf x}_0^{(k)})\phi_j({\bf x}_{\min(\tau,\tau^*)}^{(k)})$, where $N$ is the number of trajectories \cite{dinner2019galerkin,dinner2021koopman}. Importantly, the distribution $\mu({\bf x})$ is arbitrary, which allows for concentrating samples in the transition state region, where information is most needed to compute statistics such as the committor, and even allows for adaptive sampling \cite{strahan2023predicting}. In fact, theory strongly suggests that DGA is capable of remarkable data efficiency on rare-event problems when $\mu({\bf x})$ is chosen well and $\tau$ is not too long \cite{cheng2024surprising}. A further advantage is that DGA can treat microscopically irreversible processes.

When the $\phi_i({\bf x})$ are indicator functions for discrete sets of microscopic states, DGA reduces to stopped MSMs with statistic-specific boundary conditions. However, the formulation above opens the door to other choices of basis functions. Strahan et al.\ \cite{dinner2021koopman} introduced a construction multiplying a set of pairwise distances by a function that vanishes on the reactant and product state boundaries, then whitening the result through singular value decomposition (SVD). In related approaches, Khoo et al.\ \cite{khoo2025fast} and Cao and Huang \cite{cao2026continuous} consider polynomial, Fourier, and radial basis functions. Basis functions have also been learned: Thiede et al.\ \cite{dinner2019galerkin} used diffusion maps, and Strahan et al.\ \cite{strahan2023inexact} used a subspace iteration to train neural networks. 

DGA has now been applied broadly. Applications include computing committors, mean first-passage times, reactive fluxes, and equilibrium probabilities for protein folding (WW, trpcage, NTL9) \cite{dinner2019galerkin,dinner2021koopman,strahan2024bad}, fold switching (KaiB, AIB$_9$) \cite{zhang2024temperature,strahan2023inexact}, complex dissociation (insulin hexamer, dimer) \cite{antoszewski2021kinetics,antoszewski2020insulin,jeong2024analysis}, and voltage sensing (Ci-VSP) \cite{guo2024dynamics}, as well as transitions in atmospheric models \cite{finkel2021learning,finkel2023revealing,finkel2023data}. Committors accurate to within 10\% of reference values can be computed \cite{lorpaiboon2024accurate}. The availability of accurate committors for millions of sampled states makes it possible to use the committor as a reaction coordinate and to plot potentials of mean force (PMFs) as a function of it. In particular, for Ci-VSP, this analysis revealed intermediates hidden from displacement charges, reconciling electrophysiological and fluorescence measurements \cite{guo2024dynamics,villalba-galea2008s4-based}. These committors also enable computing reactive fluxes that reveal transition pathways, their relative importance, and the exchange among them \cite{dinner2021koopman,antoszewski2021kinetics,guo2024dynamics,jeong2024analysis}.

Several strategies extend DGA beyond the Markov assumption to improve accuracy. Within the DGA framework, delay embedding offers one route \cite{dinner2019galerkin,dinner2021koopman}. Memory can also be incorporated through a Mori-Zwanzig formalism \cite{lorpaiboon2024accurate}, similar to quasi-MSMs \cite{huang2020memory,huang2023memory}; doing so appears to decrease the amount of data needed to achieve a desired precision \cite{lorpaiboon2024accurate}. DGA can also estimate statistics on the fly to accelerate convergence in trajectory-stratification-based enhanced sampling \cite{dinner2009separating,dickson2010enhanced,dinner2018trajectory,strahan2024bad,vanden2009tilting}. Open questions include how best to sample training trajectories, how to learn basis functions robustly, and how to extend the approach to a formulation of TPT that accommodates history-dependent statistics \cite{lorpaiboon2022augmented}.

\newpage
\section{Perspective \thesection: From Projection Operators to Markov State Models: Features, Transition States, and Memory}
\authorblock{Michael S. O'Connor and Xuhui Huang*}

\textbf{From Projection Operators to MSMs:} From the perspective of statistical mechanics, a MSM can be viewed as a projection of the full conformational dynamics onto discrete states~\cite{Wang2018-eu}. In the Mori–Zwanzig projection operator formalism, this projection removes degrees of freedom from the underlying dynamics, resulting in memory effect. The Markov assumption underlying MSMs is therefore valid only when the lag time used to construct the model is sufficiently longer than the relaxation time of these memory kernels, so that the projected dynamics can be approximated by Markovian transitions among a finite set of states. \par
This projection perspective provides a useful framework for understanding both the strengths and limitations of MSMs. Projection onto a reduced set of states inevitably introduces memory effects. Traditional MSMs address this issue by assuming that the memory kernels have largely decayed over a sufficiently long lag time, whereas Generalized Master Equation (GME) frameworks extend MSMs by explicitly incorporating memory kernels when they remain significant after projection~\cite{huang2020memory,Cao2023-bg,huang2023memory}. Complementary to explicitly modeling memory, developing state representations that better preserve the slowest dynamical modes can shorten the memory kernels introduced by coarse-graining and improve the Markovian approximation. Another limitation of conventional MSM workflows is that transition-state structural ensembles are typically not identified explicitly during model construction. Such ensembles play an important role because they define the kinetic boundaries between metastable states and govern the rates of conformational transitions. Ultimately, the goal is to construct mechanistically interpretable MSMs that provide insight into biomolecular function.\par

\textbf{Our MSM workflow} begins with biologically motivated structural descriptors, such as inter-residue distances and dihedral angles. To identify informative features, we combine two complementary strategies. Molecular Systems Automated Identification of Cooperativity (MoSAIC)~\cite{Diez2022-gd} identifies cooperative collective motions by grouping correlated structural features, while Spectral Accelerated Sequential Incoherent Selection (Spectral-oASIS)~\cite{litzinger2018rapid} selects features that best preserve the slow dynamical modes according to the variational principle of conformational dynamics. We find these approaches complementary because functionally relevant conformational changes are often collective motions involving coordinated structural rearrangements, whereas the slowest dynamical modes may sometimes correspond to localized processes that are not directly related to function. Nevertheless, automated feature selection should complement rather than replace physical intuition. Prior knowledge of the conformational change of interest remains invaluable for identifying relevant structural descriptors and for interpreting the resulting dynamical models. The selected features are subsequently analyzed using tICA implied-timescale analysis to identify persistent spectral gaps and guide the choice of the number of metastable states. \par
Given the selected features and the estimated number of metastable states, we then train TS-DAR~\cite{Liu2025-vc,Goonetilleke2025-ap}, a deep-learning framework that identifies metastable and transition-state ensembles simultaneously within a hyperspherical latent space. Unlike conventional MSM workflows that separate dimensionality reduction and clustering~\cite{Konovalov2021-hc}, TS-DAR directly learns macrostate assignments while also highlighting candidate transition-state ensembles through out-of-distribution (OOD) conformations. The resulting state assignments are used to construct an MSM, which is subsequently validated using implied-timescale convergence and Chapman–Kolmogorov tests. As with other deep-learning approaches, care must be taken to avoid overfitting. In our experience, TS-DAR models whose slowest implied timescales exceed those of a conventional tICA-MSM by an order of magnitude or more frequently exhibit overfitting and disconnected states. Candidate transition state structures are further validated by short unbiased simulations, with approximately equal commitment probabilities to neighboring metastable states indicating transition-state behavior~\cite{Lao2026-mj}.\par

\textbf{Toward Better Representations and Memory-Aware Kinetics:} 
Current workflows still rely heavily on handcrafted molecular descriptors. 
We anticipate that equivariant graph neural networks (GNNs) and foundation models 
will increasingly enable end-to-end learning directly from atomic coordinates. 
\par
From the projection-operator perspective, memory effects naturally arise from coarse-graining and therefore, in general, cannot be eliminated completely. We anticipate that future kinetic models will increasingly incorporate memory explicitly rather than treating it solely as an error source. In particular, GME and Generalized Langevin Equation (GLE) frameworks provide systematic extensions of MSMs to improve the description of long-timescale biomolecular dynamics.\cite{huang2023memory} We therefore view the future of MSMs as a progression toward improved state representations and explicit incorporation of memory effects.

\newpage
\section{Perspective \thesection: From Operator Learning to Molecular Modeling}
\authorblock{Bettina G.~Keller*}
\newcommand{\bk}[1]{{\textcolor{blue}{#1}}}
%
MD trajectories are high-dimensional, stochastic, and often impossible to interpret directly. 
MSMs \cite{schutte1999direct, swope2004msm-one, chodera2007automatic, muff2008kinetic, buchete2008coarse, keller2010comparing, prinz2011markov} recast this complexity as a linear-algebra problem via the following sequence of transformation:
\textsl{(1)} The simulation encodes the system in terms of the potential energy function and the equation of motions.
\textsl{(2)} On the level of the ensemble, the equations of motion give rise to an evolution equation 
$\partial_t \rho_t = \mathcal{L}^{\dagger}\rho_t$ for the probability density $\rho_t$, which \textsl{(3)} induces the transfer operator $\mathcal{T}(\tau)$ \cite{prinz2011markov}.  
\textsl{(4)} Even though $\mathcal{T}(\tau)$ is an abstract mathematical object, its dominant eigenvalues and eigenfunctions encode a wealth of information on the molecular system \cite{prinz2011probing, keller2012markov}.
%
%

%
MSMs approximate $\mathcal{T}(\tau)$ by a matrix $\mathbf{T}(\tau)$ and thus give direct access to the dominant eigenspace.
The defining feature of MSMs is not the matrix representation itself, but the fact that the matrix elements are estimated from trajectory data, which already encode the underlying molecular dynamics.
In this way, one can bypass the evolution equation.
In this sense, MSMs are a statistical learning technique \cite{klus2018data}: they directly extract the target quantity (dominant  eigenspace) from noisy data (MD trajectories).
MSM methods therefore focus on two closely related questions: \textsl{(1)} preserving the mathematical structure of $\mathcal{T}(\tau)$ \cite{no2012variational, nuske2014variational, wu2016variational, klus2018data} and \textsl{(2)} improving the statistical efficiency of the estimators \cite{prinz2011optimal, trendelkamp2015estimation, wu2014statistically}. 
This is an exceptionally powerful paradigm.
Because the evolution equation is not formulated, comparatively little system-specific modeling is required. 
Once implemented, the same methodology can be applied across a wide range of molecular systems \cite{bowman2012equilibrium, pinamonti2018mechanism, schulz2018collective, qaisrani2026bridging} and even to experimental data \cite{keller2014complex,manz2017single}. 
However, the lack of consensus on MSM best practices \cite{suarez2020what, nagel2023toward, mey2024optimize} after more than two decades suggests that statistical learning alone may not resolve all remaining challenges.
In fact, progress in molecular modeling often comes from incorporating molecule-specific constraints. 
However, in current MSMs methods the connection to the underlying system is too weak to guide systematic improvement, leaving practitioners to tune representations and hyperparameters instead.
\textbf{Effective dynamics and evolution equation.}
System-specific knowledge can be included via the effective evolution equation. 
It governs a probability density $\rho_t(\mathbf{q})$ on the chosen state space $\mathbf{q} \in \Omega$ - and it changes if the state space is modified.
Most MSMs are formulated in collective variable (CV) spaces derived from configurations. 
This implicitly assumes that the effective dynamics can be described by a Smoluchowski-type equation with negligible memory.
However, projecting molecular dynamics onto a CV space generally introduces memory effects. 
Recent approaches explicitly account for this by replacing the Markovian evolution equation with generalized equations containing memory kernels \cite{suarez2016accurate, vroylandt2022likelihood, cao2020advantages, meyer2021numerical, dalton2025memory}.
The resulting memory terms provide valuable information about the quality of the CVs.
Similarly, analyzing the effects of nonlinear CVs on the effective dynamics might be insightful. Even when the projected dynamics admit a closed-form evolution equation, the resulting drift and diffusion terms can be highly nontrivial \cite{zhang2016effective, nateghi2025consistent} and their effect on the MSM quality is not fully understood.

\textbf{Molecular properties and ansatz function.}
Once a state space has been chosen, the remaining approximation error is determined by the difference between the dominant eigenspace and the space of ansatz functions. 
Since molecular structure constrains the dominant eigenfunctions, ansatz functions are the natural place to incorporate molecular knowledge.
Core-set models \cite{buchete2008coarse, schtte2011markov} exploit that dominant eigenfunctions are approximately constant within metastable regions and monotonously vary in the transition regions between them.
By identifying metastable core sets \cite{keller2010comparing, lemke2016density} before the operator is estimated, one introduces chemically meaningful structure into the ansatz space.
Alternatively, diffusion maps \cite{thiede2019galerkin} and neural networks \cite{mardt2017vampnets, rabben2020isokann, debnath2026random} learn system-specific ansatz functions directly from trajectory data, whereas chemically motivated ansatz functions \cite{vitalini2015basis} encode prior knowledge explicitly.

\textbf{Statistical uncertainty.}
In MSMs, as in any statistical learning technique, statistical uncertainty is a central challenge. 
Dynamic reweighting methods \cite{keller2024dynamical, mey2014xtram, wu2016multiensemble, stelzl2017dynamic,  donati2018girsanov, zhang2026pi} address this challenge by estimating transition probabilities from biased simulations, thereby reducing the statistical uncertainty on rare transitions.

A more radical alternative is to abandon the purely data-driven route altogether and discretize the evolution equation directly.
In the square-root approximation (SqRA), transition probabilities are obtained analytically from the discretization grid and the local (free-)energy landscape \cite{bicout1998electron, donati2018estimation, donati2021markov, donati2022review}. 
Because the matrix elements are not estimated from trajectory counts, the resulting models are free of statistical uncertainty. 
Although limited to low-dimensional spaces, SqRA can serve both as a source of sampling-free reference models and as a bridge between CV spaces, such as the translational-rotational space of two molecules \cite{zupan2025toward}. 
In conclusion, MSM construction is not only an operator-learning problem but also a molecular modeling problem. The effective evolution equation and molecular constraints deserve as much attention as statistical estimation.

\newpage

\section{Perspective \thesection: Koopman Modeling of Molecular Kinetics}
\authorblock{Antonia S.J.S. Mey*, Feliks Nüske, Stefan Klus}

\noindent\textbf{Operator-theoretic Foundations}
Koopman operator theory~\cite{koopmanHamiltonianSystemsTransformation1931b,LaMa94,mezicSpectralPropertiesDynamical2005b} provides a generalization of the classical MSM approach by allowing more general discretizations of the underlying Koopman operator $\mathcal{K}^\tau$, which propagates the expectation of an observable $\varphi$ over time $\tau$. The underlying dynamics is often thought to be metastable and one can usually determine a number of $M$ slow relaxation processes based on a spectral gap of the operator~\cite{daviesMetastableStatesSymmetric1982b}, with their molecular mechanistic interpretation provided by the corresponding eigenfunctions~\cite{schutteDirectApproachConformational1999b,deuflhardRobustPerronCluster2005b}.

\begin{figure}[!htbp]
  \centering
  \includegraphics[width=0.78\linewidth]{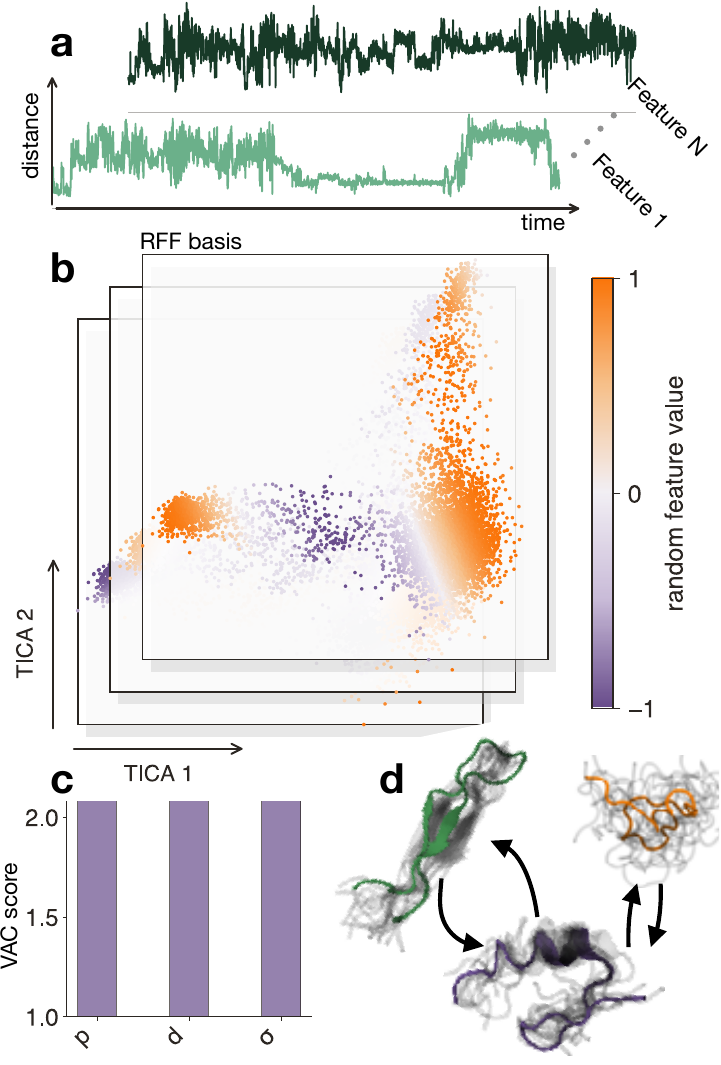}
  \caption{\textbf{Analysis of MD data with RFF}
  (a) Featurized MD trajectory.
  (b) tICA projections with random Fourier features (RFF) that make up an RFF basis.
  (c) VAC score dependency on RFF dimensions, tICA dimensions and kernel bandwidth.
  (d) Kinetic analysis showing unfolded (orange), hairpin (green), and folded state of BBA.}
  \label{fig:BBA}
\end{figure}

\noindent\textbf{Variational Approach and VAC Score}
Since the Koopman operator can in general not be computed directly, it is typically approximated from data. The variational approach to conformational dynamics (VAC)~\cite{noeVariationalApproachModeling2013b,nuskeVariationalApproachMolecular2014b} scores orthonormal candidate functions $f_i$ by $\mathrm{VAC}_M(f)=\sum_{i=1}^{M}\langle f_i,\mathcal{K}^\tau f_i\rangle_\mu$, which is maximized at $\sum_{i=1}^M\lambda_i$ by the true leading $M$ eigenfunctions (see Ref.~\cite{wu2020variational} for a generalization beyond systems obeying detailed balance). Expanding the $f_i$ in a dictionary $\Psi(x)=[\psi_1(x),\dots,\psi_n(x)]^\top$ turns this into a generalized eigenvalue problem over empirical covariance matrices estimated from time-lagged data pairs $\big\{(x_k,y_k)\big\}_{k=1}^m$. The score $\mathrm{VAC}_M$ then serves as a cross-validated metric for tuning model parameters (Fig.~\ref{fig:BBA})

\noindent\textbf{Dictionary Selection}
Several popular dictionary examples are:
\begin{itemize}[leftmargin=1ex, itemsep=0ex, topsep=0.5ex]
  \item \emph{Indicator functions} on a geometric clustering recover the
        classic MSM~\cite{schutteDirectApproachConformational1999b,klus2015numerical}.
  \item \emph{Deep learning}: the output nodes of feedforward neural networks serve as
        $\psi_i$~\cite{mardtVAMPnetsDeepLearning2018b,chenNonlinearDiscoverySlow2019b}.
  \item \emph{Kernel EDMD}~\cite{williamsKernelBasedMethodDataDriven2015b,schwantesModelingMolecularKinetics2015b}:
        $\psi_i(x)=k(x_i,x)$ gives a rich nonlinear dictionary, but costs increase cubically with the data size.
  \item \emph{Randomized dictionaries} cut this cost, e.g.\ landmark kernel
        EDMD~\cite{harriganLandmarkKernelTICA2017,litzinger2018rapid} and randomized
        networks~\cite{tabishHowDeepYour2025}. Random Fourier features~\cite{rahimiRandomFeaturesLargescale2007b}
        approximate the kernel with cosine waves
        $\psi_{\omega,b}(x)=\sqrt{2}\cos(\omega^\top x+b)$~\cite{nuskeEfficientApproximationMolecular2023b}.
\end{itemize}
Discretization-free approaches lead to higher accuracy and better generalization properties than traditional MSMs. Randomized approximations provide a compromise between generalization properties and computational efficiency and are a promising strategy~\cite{sprink_accurate_2026}. Python implementations for Koopman modeling include \href{https://deeptime-ml.github.io/}{deeptime} and \href{https://github.com/Machine-Learning-Dynamical-Systems/kooplearn}{kooplearn}.

\noindent\textbf{Molecular Kinetics from MD Data with RFF}
Figure~\ref{fig:BBA} shows an example workflow using BBA (PDB: 1FME)~\cite{lindorff-larsen2011how}. Initial featurized trajectories (e.g., heavy atom distances, Fig.~\ref{fig:BBA}a) are projected into tICA space, and RFFs are applied to the tICA space to create an RFF basis, which then serves as the input for model construction (Fig.~\ref{fig:BBA}b). Since this method has some parameter choices, such as kernel bandwidth and number of features, checking how these modeling choices change the VAC score can give confidence in the model; here Fig.~\ref{fig:BBA}c shows that a range of RFF feature sizes ($p$), tICA dimensions ($d$) and kernel bandwidths ($\sigma$) yield similar scores.  The approximated eigenpairs then allow a classic MSM analysis, recovering unfolded, folded, and beta-hairpin states (Fig.~\ref{fig:BBA}d).

\noindent\textbf{Beyond Black-Box Modeling: Generator Learning}
A related approach learns the \emph{generator} $\mathcal{L}$ rather than the finite-lag operator $\mathcal{K}^\tau$ (generator EDMD, gEDMD~\cite{klus_data-driven_2020})\footnote{\href{https://github.com/sklus/d3s}{d3s} provides a software implementation.}. The generator captures the dynamics instantaneously instead of over a fixed lag time $\tau$. This removes the lag-time choice and, more importantly, allows kinetics to be recovered even from biased or non-equilibrium simulations~\cite{devergneBiasedUnbiasedDynamics2024b,moqvistThermodynamicInterpolationGenerative2025c}. It also links naturally to coarse-grained models~\cite{zhangEffectiveDynamicsGiven2016b}, offering a data-driven route to reaction coordinates and reduced dynamics~\cite{nateghi2024kinetically,nateghi_consistent_2026}. 

\begin{figure*}[b]
\includegraphics[width=0.85\linewidth]{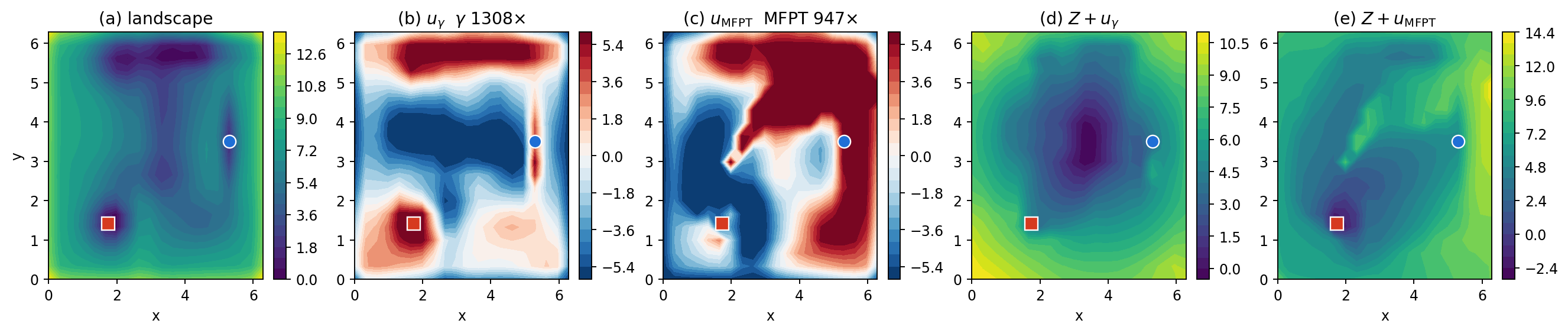}\vspace{-1mm}
\captionof{figure}{Nine-well example at \(U_{\max}=6\,\kBT\). Blue circle: source \(A\); red square: target \(B\). (a) Unbiased landscape. (b) Spectral optimum \(u_\gamma\) gives \(\gamma/\gamma_0=1308\) but is nearly symmetric. (c) MFPT optimum \(u_{\rm MFPT}\) gives \(\tau^0_{AB}/\tau_{AB}=947\). (d,e) effective landscapes \(Z+u\).}\label{fig:collapse}
\end{figure*}

\newpage
\section{Perspective \thesection: Objective choice controls optimal biasing on MSM networks}
\authorblock{Edina Rosta* and Ahmed Fadhluddin}

\textbf{Network model.}
We ask a design question: given an ordinary MSM built from unbiased trajectories, which static per-state bias \(\bd\) best \emph{accelerates} a chosen sampling objective?  An MSM replaces continuous dynamics by transitions among discrete states, so all kinetic quantities depend on the state partition, lag time, and source ensemble.  We use a reversible row-generator \(\Kzero\) with stationary distribution \(\pizero\); a dimensionless state bias \(b_i=\beta u_i\) tilts every rate by a local-detailed-balance factor, raising or lowering each state's equilibrium weight~\cite{wu2014optimal}:
\begin{equation}
  \Kb_{ij}=\Kzero_{ij}\exp\!\left[-\frac{b_j-b_i}{2}\right]\quad(i\ne j),
  \qquad
  \pi^{(b)}_i\propto \pizero_i e^{-b_i},
  \label{eq:biased-network}
\end{equation}
with diagonals set by row conservation---the DHAM bias-accounting principle~\cite{rosta2015free}.  The inputs are just \(\Kzero\) and a candidate \(\bd\); no new trajectories are run---once \(\bd\) is fixed, spectra and first-passage observables follow analytically from \(\Kb\).

\textbf{Two kinetic objectives.}
For the same bias budget \(|b_i|\le U_{\max}\) which caps the energy added at any state, two one-shot optimization problems are
\begin{equation}
  \bg=\arg\max_{|b_i|\le U_{\max}} \gamma(\Kb),
  \qquad
  \bmfpt=\arg\min_{|b_i|\le U_{\max}} \tauAB(\Kb),
  \label{eq:objectives}
\end{equation}
where \(\gamma=-\lambda_2\) is the spectral gap and \(\tauAB\) is the MFPT from source set \(A\) to target set \(B\).  The first objective accelerates global decorrelation; the second accelerates one specified transition.  Our recent MFPT-based method introduced direct optimization of this second quantity~\cite{wei2024kinetics}---to our knowledge the first enhanced-sampling bias whose objective is an explicit kinetic observable rather than an equilibrium, histogram, or variance target.

\textbf{MFPT is a conditioned observable.}
For a generator, the MFPT is an absorbing-boundary observable, not a generic steady-state arrival fraction; the source distribution and target boundary must be specified explicitly~\cite{kells2020correlation}.  Given a source distribution \(\mu_A\) on \(A\), let \(N\) index the non-target states, \(M=\Kb_{NN}\) the sub-generator (\(\Kb\) restricted to \(N\)), and \(m_N\) the vector whose entry \(m_i\) is the MFPT from state \(i\) to \(B\).  The absorbing-boundary equations are
\begin{equation}
  M m_N=-\bm 1,\qquad m_B=0,\qquad \tauAB=\mu_A^{\mathsf T}m_N .
  \label{eq:mfpt-linear}
\end{equation}
so \(\tauAB\) is the source-averaged first-passage time.
A single adjoint solve, \(M^{\mathsf T}\ell=\mu_A\), gives
\begin{equation}
  \partial_{b_i}\tauAB=-\ell^{\mathsf T}\left(\partial_{b_i}M\right)m_N,
  \label{eq:mfpt-adjoint}
\end{equation}
with \(\partial_{b_i}\Kb\) known analytically from Eq.~\eqref{eq:biased-network}.  Together with the Hellmann--Feynman derivative of \(\lambda_2\), these formulas give per-state optima by box-constrained L-BFGS-B.


\vspace{-0.4cm}
\begin{figure}[h]\centering\includegraphics[width=0.82\linewidth]{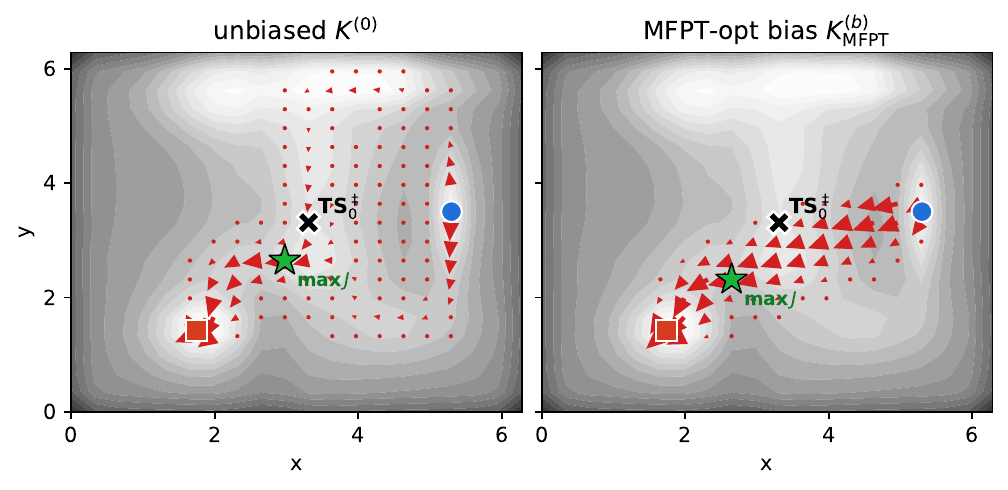} \vspace{-0.5cm}\caption{Reactive current $A\!\to\!B$: the MFPT bias amplifies the same tube; the 0\,K saddle (TS$^\ddagger_0$) lies off the flux maximum (max\,$J$).}\label{fig:rxncurrent}\end{figure} 

\vspace{-0.3cm}
\textbf{Result and take-home.}
Figure~\ref{fig:collapse} shows that the two optima are geometrically distinct.  The spectral optimum fills basins and lowers bottlenecks, treating slow transitions symmetrically; at \(U_{\max}=6\,\kBT\) it gives \(1308\times\) faster global relaxation yet marginally \emph{slows} the chosen transition (\(0.76\times\)).  The MFPT optimum is directional: it destabilises the source side while preserving the target basin, giving \(947\times\) MFPT speedup.  A bias is also judged by how faithfully it preserves the mechanism, which transition-path theory quantifies as the overlap \(\Omega_J\) of the biased and unbiased reactive currents (\(\Omega_J=1\): pathway unchanged).  On the acceleration--\(\Omega_J\) Pareto front the MFPT-optimal bias lies above the spectral bias---greater \(A\!\to\!B\) acceleration at any given overlap.  Thus optimal biasing on an MSM is objective-dependent~\cite{wang2026objdep}: spectral gaps are appropriate for global equilibration, while MFPTs are the natural target for specific kinetic acceleration.  A slow eigenmode is therefore not, by itself, a reliable surrogate for a chosen first-passage event.

\newpage
\section{Perspective \thesection: MSMs, Collective Variables, and Effective Dynamics}
\authorblock{Christof Schütte*}
\newcommand{\Tbf}{\mathbf{T}}
\newcommand{\Pcal}{\mathcal{P}^{\tau}}
\newcommand{\Kcal}{\mathcal{K}^{\tau}}
\newcommand{\xit}{\xi(\xbf_t)}

Markov state models (MSMs) are often introduced as a pragmatic way of turning long molecular trajectories into a finite-state, row-stochastic transition matrix \(\Tbf\), whose eigenvalues \(\lambda_i\), implied timescales, transition rates, and transition-path-theory (TPT) fluxes are interpreted kinetically.  
This view is useful but too narrow: MSMs are one instance of transfer-operator methods for learning slow dynamics from data.
MSM thinking goes beyond discretizing configuration space: it guides the discovery of coordinates, effective dynamics, and transition mechanisms.

\paragraph{MSMs are one ansatz in a transfer-operator family}
At lag time \(\tau\), the microscopic dynamics propagate densities by 
$ p_{t+\tau}(\xbf_{t+\tau})=\int p^{\mathrm{prop}}(\xbf_{t+\tau}|\xbf_t)\,p_t(\xbf_t)\,d\xbf_t$.
The transfer operator \(\Pcal\) is the mathematical object defined by the map \(p_t\mapsto p_{t+\tau}=\Pcal p_t\). 
Thus, the operator is an exact object of the underlying dynamics, not a numerical model.
The modeling step begins when we approximate this operator from data in a finite-dimensional ansatz space.  MSMs make one particularly transparent choice: indicator functions of sets.  This choice turns the approximation of \(\Pcal\) into linear algebra on \(\Tbf\), and it explains why MSMs are interpretable and computationally robust.
This set-wise construction is only the simplest Galerkin choice within a family of transfer-operator approximations governed by variational principles, spectral approximation, and data-driven model reduction \cite{SchuetteKlusHartmann2023,KlusNueskeKoltaiWuKevrekidisSchuetteNoe2018}.  Once the ansatz space is no longer restricted to set indicators, one may use problem-adapted functions: diffusion coordinates, extended dynamic mode decompositions with learned dictionaries, time-lagged autoencoders, VAMPnets, ISOKANN-type eigenfunction learning, or neural solutions of eigenvalues of $\Pcal$ \cite{LDBK17,AutoencoderNoe2018,MardtPasqualiWuNoe2018,sikorski2024learning,ZhangLiSchuette2022EigenvaluePDEs}.  Thus, modern neural architectures do not replace MSMs.  They generalize the MSM idea from a partition-based basis to nonlinear, trainable ansatz functions that can resolve slow modes in high dimension and adapt their resolution to accuracy demands.

\paragraph{Collective variables that preserve relevant dynamics}
The next step is not merely to improve MSMs, but to ask whether there is a dynamically-informative collective variable (CV) or reaction coordinate that maps the high-dimensional states $\xbf$ of the system to a much lower dimensional CV-space with reduced coordinates  \(\xi(\xbf)\).  A CV is dynamically-informative when it forms a coordinate system of the transition manifold of the system \cite{Bittracher2018}. In this case, the effective process in CV-space accurately reproduces the dominant relaxation timescales and transition probabilities of the full system.  The optimal-CV formulation of \cite{ZhangSchuette2025} makes this criterion explicit by connecting it to the transfer operator and by minimizing the deviation between full and effective dynamics.  This shifts the target from ``find a physically intuitive reaction coordinate'' to ``find a low-dimensional CV whose effective dynamics are faithful for the slow modes''.  The CV is then not merely a heuristic reaction coordinate; it is a learned interface between the high-dimensional microscopic process and a reduced process that carries the same dominant kinetic information.
Earlier work on effective dynamics, that is the reduced dynamics in CV-space, provides complementary evidence that this reduced dynamics can be both mathematically controlled and computationally learned \cite{ZhangHartmannSchuette2016,ZhangSchuette2017,LelievreZhang2019,ZhangSchuette2023Understanding,WuNoe2024}.  This point is important for the interpretation of MSMs.  A large transition matrix on a fine partition may be an intermediate construction, while the real scientific object of interest is a dynamically-informative CV together with their effective transfer operator.  Thus the best MSM output may be not only a large matrix, but also a smaller and more predictive description of the system.

\paragraph{CVs enable mechanistic insight, not only kinetics}
Once such CVs and their effective dynamics are available, the analysis can go beyond estimating rates.  TPT, committors, reactive currents, flux decompositions, and transition-path ensembles become low-dimensional diagnostic tools for explaining how transitions occur \cite{MetznerSchuetteVandenEijnden2006TPTexamples,BerezhkovskiiSzabo2019CommittorsMilestones}.  In the full space these quantities may be difficult to visualize, unstable to estimate, or hard to communicate.  In an optimal or dynamics-informative CV space, however, they can reveal where probability accumulates, where reactive trajectories cross bottlenecks, which channels dominate the flux, and which intermediate regions are kinetically relevant rather than merely geometrically nearby.
Recent MSM-inspired neural CV learning and transition-pathway reconstruction shows how learned CVs can support an interpretable picture of effective dynamics and pathways \cite{Sikorskietal2025,SDWS25}.  This closes the circle.  MSMs introduce the operator viewpoint in a concrete and accessible form; transfer-operator learning broadens the class of admissible approximations; optimal CV theory identifies reduced variables that preserve the important dynamics; and TPT turns the resulting reduced dynamics into mechanistic explanations.  

\vspace{-0.2cm}
\paragraph{Summary}
MSMs are  not an endpoint.  They are a gateway from raw trajectory data to operators, from operators to optimal coordinates, and from coordinates to mechanistic narratives about metastability, pathways, and rare events.

\newpage
\newpage
\section{Perspective \thesection: Feature selection via MoSAIC contact clusters}
\authorblock{Gerhard Stock*}
The selection of suitable input coordinates (or features) is arguably
the most important step in constructing a MSM \cite{scherer2019variational,
konovalov2021markov, nagel2023selecting}. Since Cartesian atom
coordinates may be flawed \cite{sittel2014principal}, dihedral angles
mainly capture local dynamics \cite{sittel2017principal}, and
$C_\alpha$-distances scale quadratically with system
size \cite{nagel2023selecting}, we favor inter-residue contact
distances \cite{latzer2008conformational, 
ernst2015contact, yao2019establishing}. Reflecting short-range
interactions, they directly reveal microscopic mechanisms while
capturing long-range structural changes. To improve microstate
clustering and implied timescales, we apply Gaussian low-pass
filtering to the distance trajectories \cite{sartore2026lost}.
Because only a small subset of these coordinates is involved in a
given biomolecular process, we reduce their number
\cite{litzinger2018rapid, sittel2018perspective, brandt2018machine,
  ravindra2020automatic, suarez2020what, glielmo2021unsupervised,
  ahalawat2023resolving, diez2022correlationbased}. Using the community
detection--based method MoSAIC \cite{diez2022correlationbased}, we
identify correlated dynamics and partition proteins into dynamically
connected contact clusters. We illustrate this approach for three
representative systems: T4 lysozyme
(T4L) \cite{dixon1992structure,hub2009detection, lange2008full,
  ernst2017identification, post2022cooperative, nagel2024accurate,
  diez2025recovering}, the villin headpiece
(HP35) \cite{kubelka2008chemical, piana2012protein,
  nagel2019dynamical, nagel2023benchmark}, and the PDZ3
domain \cite{petit2009hidden, bozovic2021speed, janke2025universal,
  ali2022nonequilibrium, ali2024allosteric, dorbath2026contact}.

\begin{figure}[ht!]
\centering
\includegraphics[width=.4\textwidth]{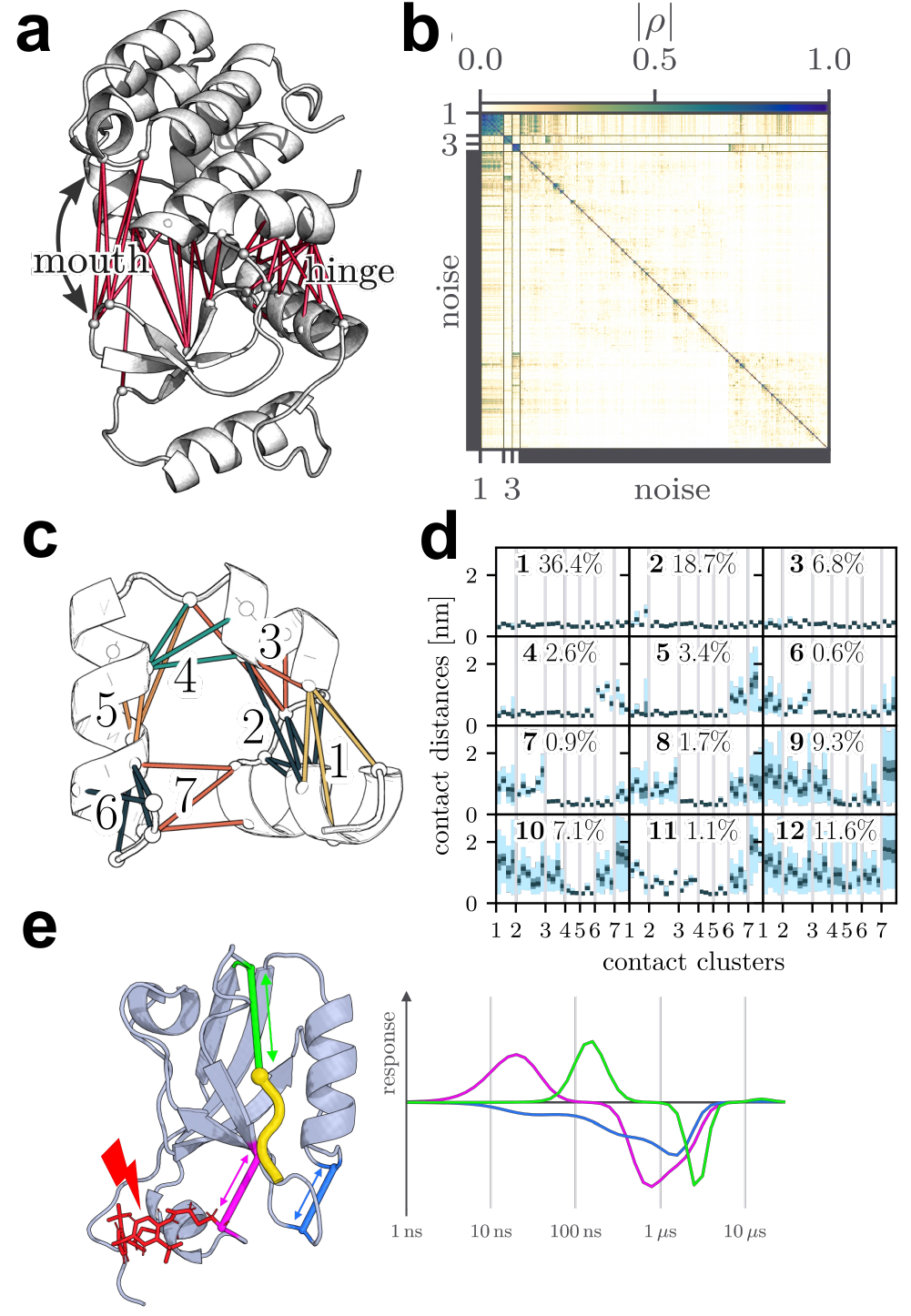}
\caption{
T4L: (a) Structure and main contact cluster; (b) MoSAIC correlation
matrix. HP35: (c) Structure and contact clusters; (d) contact
distance distribution of the 12 metastable states. (e) PDZ3:
Structure and representative contacts including timescale
analyzes. Figures were adapted from
Refs.\,\cite{nagel2024accurate, nagel2023selecting, ali2024allosteric}}
\label{fig:stock}
\end{figure}

To elucidate the open-closed transition of T4L (Fig.\,\ref{fig:stock}a), we
analyzed $\sim\!400$ inter-residue contacts from a $60\,\mu$s-long MD
simulation \cite{ernst2017identification}, using
MoSAIC \cite{diez2022correlationbased}. The block-diagonal correlation
matrix in Fig.\,\ref{fig:stock}b shows that about 90\,\% of the coordinates
are uncorrelated and can be regarded as noise. Three small clusters
describe local motions, whereas one dominant cluster comprising
$\sim\!30$ contacts captures the functional
dynamics \cite{post2022cooperative}. Extending from the hinge to the
mouth region, these contacts mediate the cooperative transition: local
changes in the flexible hinge propagate through the relatively rigid
backbone, driving opening and closing of the
mouth \cite{ernst2017identification}. Using five essential contacts as
coordinates \cite{brandt2018machine}, we constructed an MSM that
reveals the heterogeneity of the transition
paths \cite{post2022cooperative}. 

The folding of HP35 was studied in a publicly available
$300\,\mu$s-long MD simulation by Piana et al. \cite{piana2012protein}.
To establish a benchmark MSM, we compared feature sets, dimensionality
reduction methods, and clustering schemes \cite{nagel2023selecting,
  nagel2023benchmark}. Native contact distances outperformed all
alternatives. Combined with the MoSAIC contact classification, they
provide a concise structural characterization of the 12 metastable
states (Figs.\,\ref{fig:stock}c,d): the first five states form the native
basin, states 6--8 are intermediates, and states 9--12 constitute the
unfolded basin. Although alternative feature sets reproduced this
overall picture, they often failed to resolve the experimentally
observed substates of the native basin \cite{kubelka2008chemical}, and
inadequately described the initial folding step involving contacts
from clusters 4 and 5. 

PDZ3 is a minimal model of allosteric communication, in which
perturbation of the short C-terminal $\alpha_3$-helix modulates ligand
binding affinity (Fig.\,\ref{fig:stock}e) \cite{petit2009hidden}.
Time-resolved infrared spectroscopy showed that photoswitching of
$\alpha_3$ triggers conformational changes across the protein on
timescales from 0.1\,ns to tens of $\mu$s \cite{bozovic2021speed,
  janke2025universal}. Extensive nonequilibrium MD simulations combined
with MoSAIC reproduced the experimental timescales and explained them
as a multistep process involving cooperative contact changes within
clusters and communication between distant
clusters \cite{ali2022nonequilibrium, ali2024allosteric,
  dorbath2026contact}. Figure\,\ref{fig:stock}e compares the timescales of
initially excited (purple) and distant (green) contacts, illustrating
the allosteric response.

In summary, MoSAIC clustering of contact distances provides an
effective feature set for Markov modeling, enables structural
characterization of metastable states, and reveals the mechanisms
underlying biomolecular processes.

\newpage
\section{Perspective \thesection: VAMP is not a panacea}
\authorblock{Vincent Voelz*}

MSMs \cite{chodera2006long,noe2008transition,prinz2011markov,bowman2013introduction} and related master equation approaches \cite{buchete2008coarse} came to widespread attention in the 2000s as a tool to understand protein folding. MSMs gave the community a tantalizing set of tools to analyze ensembles of non-equilibrium trajectories generated using distributed computing \cite{noe2009constructing,voelz2010molecular}: a "network" view of the complete folding landscape \cite{rao2004protein,bowman2010protein} , a framework for understanding how microscopic kinetics give rise to macroscopic folding phenomena \cite{bowman2011taming,beauchamp2012simple,voelz2012slow}, and a toolkit for calculating rates and pathways \cite{metzner2009transition,noe2009constructing}. In the following decade, the powerful variational principle known as VAMP (Variational Approach to Markov Processes) was developed to select models that optimally capture molecular kinetics \cite{noe2013variational,wu2020variational}.  This made possible unsupervised machine learning pipelines for optimal MSM construction \cite{scherer2019variational}, including those based on neural networks (e.g. VAMPnets \cite{mardt2018vampnets}). Today, deep learning is enabling the development of new methods that can identify slow collective variables \cite{mehdi2024enhanced,cao2025amuset}  resulting in a new generation of MSM-based approaches:  generative molecular dynamics models \cite{wu2018deep,jing2024generative},   GraphVAMPnets \cite{liu2023graphvampnets}, State-free Reversible VAMPnets \cite{chen2019nonlinear,sidky2019high},   non-Markovian methods \cite{liu2025memory,dominic2023building}, and methods to reconcile MSMs with ensemble-averaged experimental measurements \cite{raddi2025model,nguyen2025high} or dynamical observables \cite{kolloff2023rescuing}.  

\paragraph{Constructing MSMs remains an art}
While VAMP-based pipelines are designed to construct MSMs with minimal supervision, several practical issues should be considered.

1. The slowest, most time-correlated motions are often not the ones of interest. This is a problem we continue to encounter  analyzing trajectories from Folding@home \cite{voelz2023folding}: massively parallel simulation inevitably leads to outliers (e.g. a rogue unfolding trajectory among many start from a folded state).  VAMP and its linear cousin TICA \cite{schwantes2013improvements,perez2013identification} incorrectly identify these outliers as important time-correlated motions, requiring intervention via careful data curation.

2. Selection of input features and model hyper-parameters is nontrivial \cite{nagel2023benchmark}.  VAMP has made MSM construction more rigorous, but arguably more challenging: what featurizations of your data result in the best VAMP score?  Should you compare as many models as you can? When testing large numbers of possibilities, testing/training cross-validation becomes especially important to avoid overfitting and select the best models.

3. An interpretable model may be more desirable than a variationally optimal model. For example, VAMPnets and related methods give variationally optimal neural network-based state classifications, but—as is often the question with machine learning models—what physical insight is being learned?  Simpler linear feature-based approaches like TICA have enjoyed enduring popularity, in part due to their advantage in interpretability.

4. Several software packages are available to construct and analyze MSMs \cite{scherer2015pyemma,hoffmann2021deeptime,harrigan2017msmbuilder,porter2019enspara,nagel2023msmhelper}, but there is no single software package that contains all desired functionality, and some packages are not actively maintained. Each new project requires consideration of the software tools to be used.

\paragraph{Estimating equilibrium populations with MSMs remains challenging}

While MSMs make excellent estimates of rates and pathways of rare events, inferring equilibrium populations from kinetic transitions is a challenge. Consider a protein that folds in a two-state manner, having a modest folding stability of 5 kcal/mol.  The folded-to-unfolded reaction at room temperature has an equilibrium constant of $K \approx 0.0002$. Stability could be estimated from the ratio of the unfolding rate to the folding rate ($K = k_u/k_f)$, but it is easy to see the practical difficulty: even with a fast folding rate of $k_f = 10^6$ s$^{-1}$, estimating $K$ would require (on average) 5 milliseconds of aggregate trajectory data started from the folded state just to observe \textit{one} unfolding event;  precise estimates would require even more sampling. More generally, MSMs have a "chicken-and-egg" problem: MSM estimators typically assume that  observed transition counts are generated at equilibrium;  for rare event sampling, this is usually never the case.  

To better estimate equilibrium populations, two main approaches have been pursued. The first is the use of multi-ensemble Markov Models (MEMMs) \cite{wu2016multiensemble,keller2024dynamical,rosta2015free,wan2016maximum},  which take as input energy samples and transition counts from biased thermodynamic ensembles in which enhanced sampling can be performed \cite{paul2017protein,ge2022estimation,zhang2024improved}. The most notable of these methods are the TRAM estimator \cite{wu2016multiensemble}, DHAM \cite{rosta2015free}, and more recent Girsanov reweighting approaches.\cite{donati2022review,zhang2026pi}  

A second approach has focused on reweighting the transition counts and/or the trajectories used to build MSMs. Adaptive seeding studies from our group showed that reweighting counts by prior state probability estimates improves predictions of equilibrium populations \cite{wan2020adaptive}.  The Zuckerman group has improved on this idea by developing iterative reweighting approaches \cite{russo2020iterative,kania2026randomized}, which can be used both for steady-state and equilibrium conditions. These methods represent an exciting way forward towards accurate estimates of populations as well as rates.

\newpage
\section{Perspective \thesection: Two matrices to rule them all}
\authorblock{Daniel M. Zuckerman*}

\newcommand{\pinit}{p^\mathrm{init}}
\newcommand{\pprop}{p^\mathrm{prop}}
\newcommand{\tij}{T_{ij}}
\newcommand{\tmat}{\mathbf{T}}

\begin{figure}[h]
    \centering
    \includegraphics[width=0.85\linewidth]{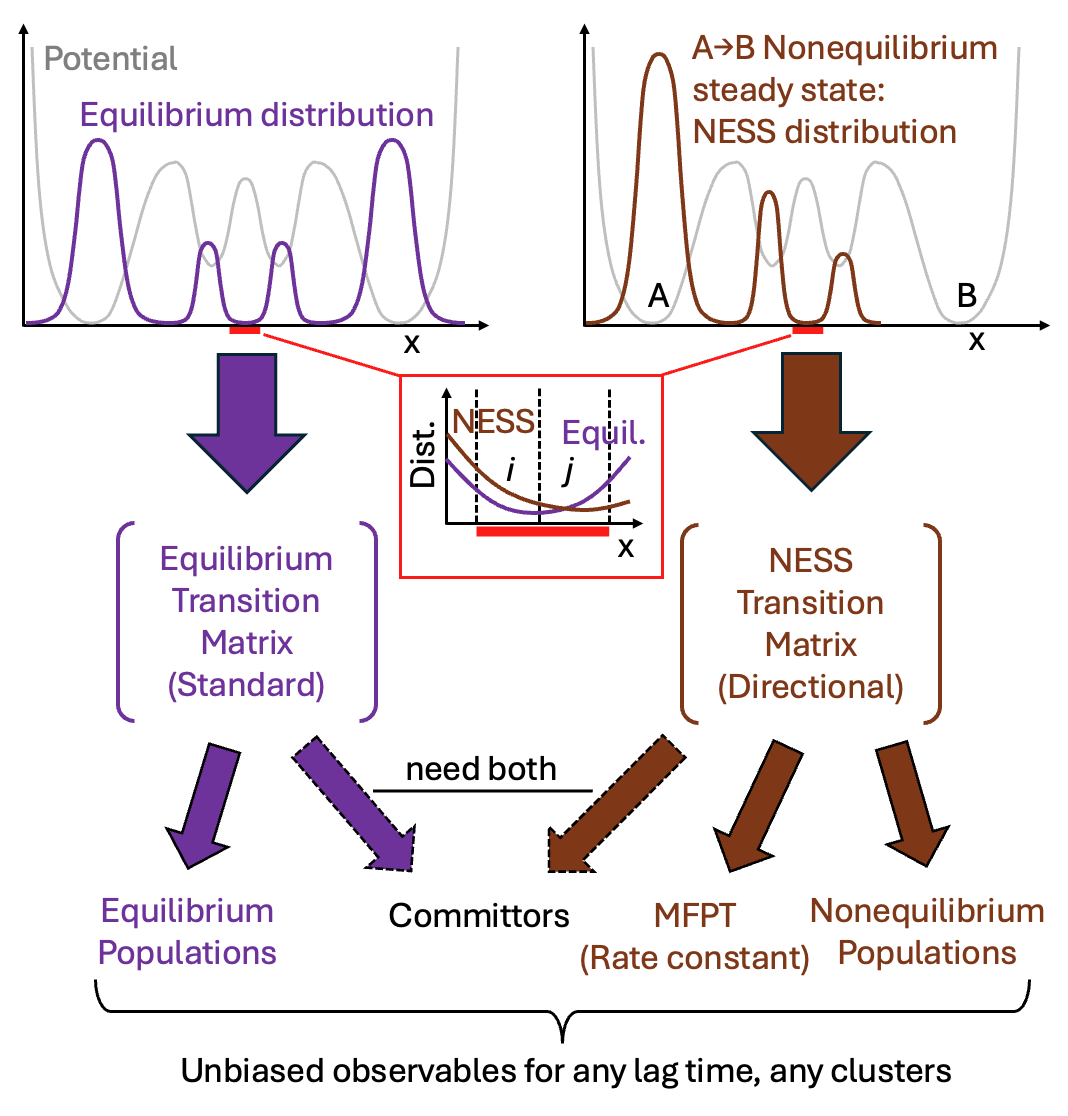}
    \caption{Unbiased observables from two transition matrices.  Schematic illustration of transition matrices derived from either equilibrium (purple) or nonequilibrium steady state (NESS; brown) data.  In the pertinent NESS, trajectories reaching sink state B are restarted in source state A.  Each matrix can generate unbiased estimates of different observables, although notably committor values require both matrices.  The inset (red box) illustrates differing within-state distributions which ultimately lead to different matrix elements via \eqref{tij-dmz}.}
    \label{fig:zuckerman}
\end{figure}

Why compromise if we don't have to?  Typical MSMs build in compromises in both space (\emph{are conformational states fine enough for specificity, yet coarse enough for sufficient statistics?}) and time (\emph{is the lag time long enough for quasi-Markovian behavior, yet short enough to resolve key mechanistic events?}) \cite{prinz2011markov}.  The net result is that MSM estimates of key observables are often biased, even with sufficient training data \cite{suarez2020what,aristoff2026msmsrevisited}.  However, such compromises are unnecessary if we are willing to abandon the conventional MSM framework relying on a single transition matrix -- and implicitly the Markov assumption itself.  Instead, we can augment the usual equilibrium-based transition matrix with a nonequilibrium, directional matrix (Fig. \ref{fig:zuckerman}) to calculate unbiased observables, i.e., quantities that will be exact in the limit of infinite data, regardless of spatial coarse-graining or lag time.

To see how all this works, we should review the MSM ``pipeline'' from Markovian microscopic dynamics to coarse-grained states \cite{prinz2011markov}.  In the full phase space of all system coordinates, $\xbf_t$ is a Markov process, propagated by transition probability $\pprop$ so that an arbitrary distribution $p_t$ at time $t$ evolves according to \cite{prinz2011markov,aristoff2026msmsrevisited}
\begin{equation}
    p_{t+\tau}(\xbf_{t+\tau}) = \int d\xbf_t \; \pprop( \xbf_{t+\tau} | \xbf_t) \; p_t(\xbf_t)
    \label{prop-dmz}
\end{equation}
From the microscopic dynamics, we coarse-grain phase space into regions $i$ (or $j$) and derive transition matrix elements $\tij$, which are conditional transition probabilities obtained from the microscopic dynamics $\pprop$ as well as the distribution of phase points available from trajectory data $\pinit$:
\begin{equation}
    \tij 
    \propto \int_i d\xbf_{t} \int_j d\xbf_{t+\tau} \; \pprop( \xbf_{t+\tau} | \xbf_t ) \; \pinit(\xbf_t) 
    \label{tij-dmz}
\end{equation}
where the $i$ and $j$ subscripts indicate regions of integration, and the normalizing factors are suppressed for clarity \cite{aristoff2026msmsrevisited}.

A key observation is that $\tij$ values \eqref{tij-dmz} generally depend on the data distribution $\pinit$, and this will affect all downstream observables computed from $\tmat$.  The sensitivity will be greater at shorter lag times \cite{suarez2020what}.

Despite the standard practice of seeking sufficiently long lag times \cite{chodera2014markov} that can even reach the $\sim$100 ns range \cite{suarez2020what}, theory dating to 2009 establishes that the MFPT (inverse rate constant) can be computed without bias for any lag time and state discretization so long as trajectory data corresponds to the directional process of interest (Fig.\ \ref{fig:zuckerman}) and not equilibrium \cite{vanden2009tilting,dinner2009separating,suarez2014simultaneous}.  More recent work has shown likewise that the committor and local fluxes can also be estimated without bias at arbitrary lag times and coarse-graining \cite{russo2021unbiased,aristoff2026msmsrevisited}.  The value of using short lag times is to avoid jumping over key mechanistic details that may occur rapidly \cite{suarez2020what}.

What goes wrong in estimating observables from a single transition matrix, and why does the second matrix rectify the estimators?  The key point is that standard estimation of observables such as the MFPT and committors (via ``first step'' relations \cite{privault2018understanding}) directly relies on the Markov assumption which is never precisely satisfied \cite{prinz2011markov,chodera2014markov}.  This bias in the estimators, it must be emphasized, is not due to limited sampling but rather to approximations of the MSM framework \cite{aristoff2026msmsrevisited}.  On the other hand, estimators for multiple observables (populations, MFPT, mechanistic fluxes between states) can be constructed that rely only on stationary relations, i.e., $\pi^\top \tmat = \pi^\top$, which do not rest on a Markov assumption despite appearances because they implicitly recapitulate the microscopic dynamics \eqref{prop-dmz} as has been shown \cite{aristoff2026msmsrevisited}.

Unbiased estimators must be combined with properly distributed trajectory data for accurate observables \cite{vitalis2019removal}.  Standard pipelines are ``at the mercy'' of data in hand, but reweighting can improve intra-cluster distributions \cite{wan2020adaptive} and thus $\pinit$ in \eqref{tij-dmz}.  The recent ``RiteWeight'' algorithm \cite{kania2026randomized} in principle can reweight data for both transition matrices (Fig.\ \ref{fig:zuckerman}).

\section{Concluding Remarks}
\authorblock{Daniel M. Zuckerman and Antonia S.J.S. Mey}
This document's purpose was to bring together different practitioners' thoughts and strategies for building, validating, and interpreting Markov models and related approaches. By design, this collection showcases differing opinions on the topic, assembling ideas on state-space resolution, on whether detailed balance should be imposed, on the role of deep learning, and on whether the Markov assumption is even needed. This wide range of both methodological development and application underlines both the maturity and continuing vibrancy of the field, and it highlights the overall utility of having a mathematically rooted framework for the analysis of trajectory data.

While there is no consensus on what constitutes an ideal MSM, several common themes emerge from the different perspectives. The initial representation of the data, through collective variables or dimensionality reduction, and the resulting state definitions strongly influence model quality. Contact distances, optimized reaction coordinates, core sets, and learned bases are all valid strategies, and the best choice may depend on the scientific question the model aims to address. Similarly, long-timescale effects are not necessarily best captured by using longer lag times; approaches based on master equations, generator modeling, or delay embeddings may perform better. One general caveat that cannot be resolved even by the best mathematical framework is the curse of sampling, a theme that appears again and again. Molecular simulation often means working with unconverged data, compounded by the difficulty of recovering equilibrium properties from rare-event kinetics within these samples. Relatedly, the most interpretable model need not be the variationally optimal one, since chemical or biological insight does not always coincide with variational optimality, which is why automation should complement rather than replace physical intuition.

We hope this Roundtable article serves as an overview of the landscape of Markov modeling, providing ideas and open questions and showcasing where some of the field's active research is focused at the time of writing. Readers are encouraged to engage with the content by adopting and combining some of these practices. They should equally feel inspired to contest others, paving the way for a future updated version of this document following the spirit of this living journal.

\section*{Author Contributions}
%
ASJSM co-wrote the introduction and concluding remarks.
DMZ co-wrote the introduction and concluding remarks.
All other authors wrote or co-wrote their respective perspectives.

For a more detailed description of author contributions,
see the GitHub issue tracking and changelog at \githubrepository.

%

For a more detailed description of contributions from the community and others, see the GitHub issue tracking and changelog at \githubrepository.

\section*{Potentially Conflicting Interests}

None declared.

\section*{Acknowledgments}
M O'Connor and X Huang thank Siqin Cao and Yichong Lao from UW-Madison for helpful discussions. 
G Stock thanks co-workers G. Diez, E. Dorbath, D. Nagel, and S. Sartore
who contributed to the results shown in Fig.\ \ref{fig:stock}.  DM Zuckerman is grateful for numerous discussions with David Aristoff, Jeremy Copperman, Gideon Simpson, and Rob Webber. C Schuette thanks co-workers W. Zhang, L. Donati, and M. Weber. E Rosta and A Fadhluddin used AI-based tools to assist with drafting, code, and editing of their contribution, and take full responsibility for its content.

\section*{Funding Information}
XH acknowledges support from the Hirschfelder Professorship Fund and the Vilas Associate Award from University of Wisconsin-Madison as well as the Research Forward Fund from the University of Wisconsin-Madison Office of the Vice Chancellor for Research with funding from the Wisconsin Alumni Research Foundation.  MO acknowledges a fellowship from the National Institutes of Health (NIGMS, T32 GM130550).
GS thanks the Deutsche Forschungsgemeinschaft (DFG) for continuous support within the framework of the Research Unit FOR 5099 “Reducing complexity of nonequilibrium” (project No.\ 431945604).  ARD acknowledges support from National Institutes of Health (NIH) Award R35 GM136381. JW acknowledges sabbatical support from the Margot and Tom Pritzker Science Foundation. DMZ appreciates support from NIH Grant GM115805.
BGK thanks the Volkswagen Foundation for funding through a Momentum grant.
GRB is grateful for support from NIGMS R35GM152085.

\section*{Author Information}
\makeorcid

\bibliography{refs_msm_roundtable}



\end{document}